\documentclass[twocolumn,preprintnumbers,amsmath,amssymb]{revtex4}
\usepackage{graphicx}

\usepackage{color}
\usepackage{amsmath}

\usepackage{dcolumn}
\usepackage{bm}
\usepackage{float}

\newcommand{\comment}[1]{}

\begin{document}

\title{Three-path atom interferometry with large momentum separation}

\author{Benjamin Plotkin-Swing, Daniel Gochnauer, Katherine E. McAlpine, Eric S. Cooper, Alan O. Jamison\footnote{Present address: Department of Physics and MIT-Harvard Center for Ultracold Atoms, Research Laboratory of Electronics,
MIT, Cambridge, Massachusetts 02139, USA}, and Subhadeep Gupta}

\affiliation{Department of Physics, University of Washington, Seattle, Washington 98195, USA}

\date{\today}
\begin{abstract}
We demonstrate the scale up of a symmetric three-path contrast interferometer to large momentum separation. The observed phase stability at separation of 112 photon recoil momenta exceeds the performance of earlier free-space interferometers. In addition to the symmetric interferometer geometry and Bose-Einstein condensate source, the robust scalability of our approach relies on the suppression of undesired diffraction phases through a careful choice of atom optics parameters. The interferometer phase evolution is quadratic with number of recoils, reaching a rate as high as $7\times10^7$ radians/s. We discuss the applicability of our method towards a new measurement of the fine-structure constant and a test of QED. 
\end{abstract}
\maketitle


The precision of atom interferometry \cite{cron09} enables applications, such as inertial sensing \cite{pete01,mcgu02,durf06,dutt16,geig11}, and tests of fundamental physics, such as the equivalence principle \cite{fray04,schl14} and quantum electrodynamics (QED) \cite{bouc11,park18}. Light-pulse interferometers, central to these endeavors, use standing-wave optical pulses as beamsplitters and mirrors, imparting momenta in units of photon momentum $\hbar k$ to the atoms. Such interferometers gain sensitivity by increasing the enclosed space-time area with momentum-boosting acceleration pulses \cite{mull09,chio11}. Phase-stable interferometers with large momentum separation are thus an overarching goal in atom interferometry. 

Path separations $n \hbar k$ with $n$ up to $102$ have been demonstrated \cite{chio11}, however interferometer phase stability \cite{footkas} was not observed due to technical noise from mirror vibrations. Vibration immunity and resultant phase stability can be recovered by operating two simultaneous interferometers in a conjugate or dual geometry \cite{chio09,chio11}. However, the operation of such interferometers has been limited to $n \leq 30$ \cite{mull08,chio11,asen17}.  While $n = 80$ has been reported in a guided-atom interferometer \cite{mcdo13}, the confining potential introduces additional systematic effects. 

All of these earlier works involved interference between two paths. Here we demonstrate large momentum separation in a three-path interferometer, an alternative geometry featuring an inherent immunity to many systematic effects \cite{gupt02,jami14}. We observe phase stability for very large momentum separation, achieving 30\% visibility at $n = 112$. The resulting interferometer phase grows quadratically with momentum, reaching a rate as high as $7\times10^7$ radians/s. Undesirable diffraction phases are theoretically and experimentally analyzed and controlled by our choice of atom-optics parameters. Our interferometer demonstrates favorable scaling for a precision measurement of the fine-structure constant $\alpha$ and test of QED. 

\begin{figure*}
\includegraphics[width=\linewidth]{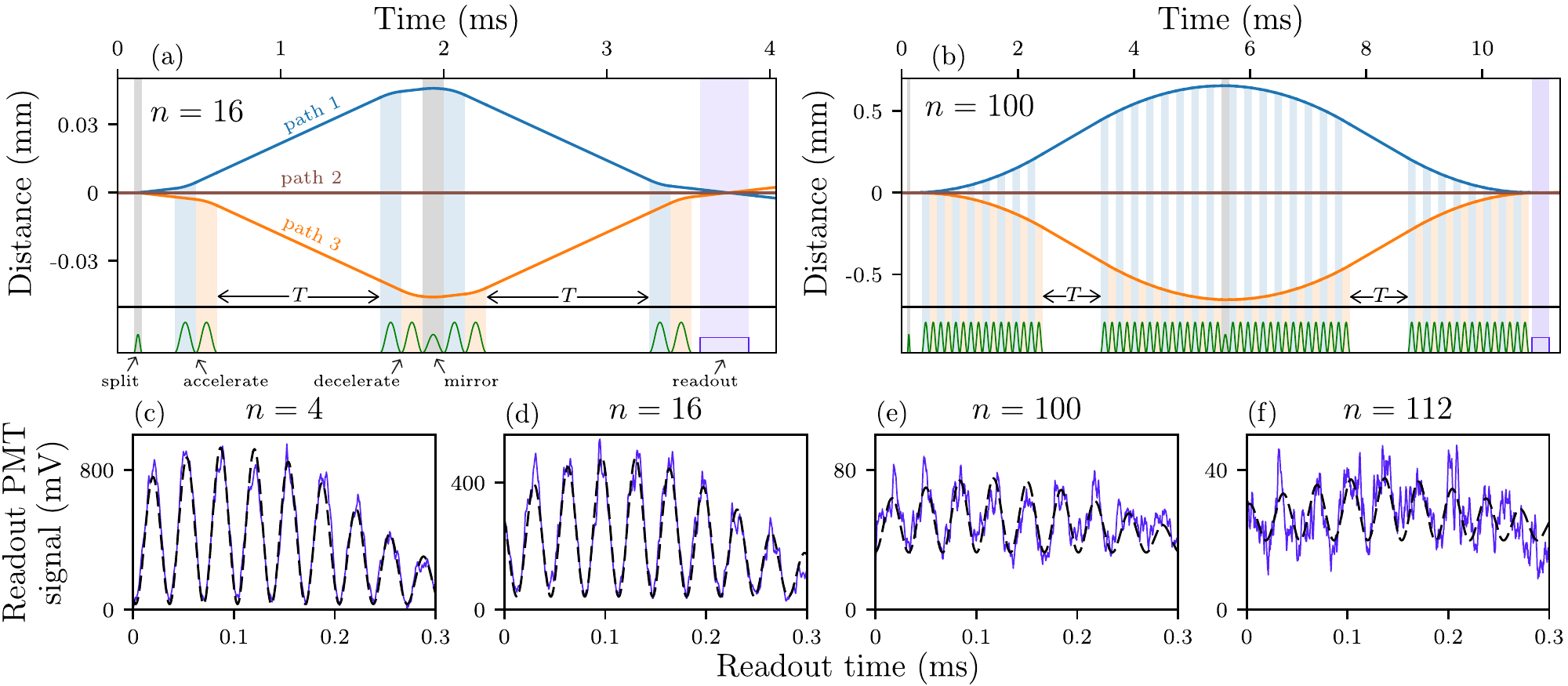}
\caption{(a,b) Space-time trajectory (to scale) and atom optics sequence for the $n=16$ and $n=100$ contrast interferometer (CI) with $T=1\,$ms. The shaded regions indicate that some pulses (black shading) affect both moving paths, while others (orange and blue shading) affect a single path. The CI signal is acquired by applying a traveling wave laser pulse (violet shading) and collecting the Bragg-reflected optical signal on a photo-multiplier tube (PMT). Readout signals (violet) with fits (dashed black) for various momentum splittings $n$ are shown in (c,d) (20 shot averages), and (e,f) (80 shot averages).}
\label{fig:fig1}
\end{figure*}

Our contrast interferometer (CI), (Fig.\,\ref{fig:fig1}) operates on a Bose-Einstein condensate (BEC) atom source and consists of four atom-optics elements: splitting pulse, mirror pulse, acceleration pulses, and readout pulse. The splitting pulse places each atom into an equal superposition of three $z$-axis momentum states: $|+2\hbar k \rangle$, $|0\hbar k\rangle$, and $|-2\hbar k \rangle$, referred to as paths 1, 2, and 3 respectively. The mirror pulse reverses the momenta of paths 1 and 3. 
The acceleration pulses increase the momentum separation of paths 1 and 3 to $n \hbar k$ during two sets of free evolution times $T$. After the final deceleration sequence brings the outer paths back to $|\pm2\hbar k\rangle$, all three paths overlap in space and form an atomic density grating with spatial period $\pi/k$, whose amplitude varies in time \cite{foot4hk}: 
\begin{equation}
\sqrt{C(t)}{\rm cos}\Big(\frac{\phi_1(t)+\phi_3(t)}{2}-\phi_2(t)\Big){\rm cos}\Big(2kz+\frac{\phi_1(t)-\phi_3(t)}{2}\Big)
\label{eq:grating}
\end{equation}
By pulsing on a traveling ``readout'' laser beam and collecting the Bragg-reflection off this matter-wave grating, we obtain its {\it contrast} as the characteristic CI signal:
\begin{equation}
S(t)=C(t){\rm cos}^2\Big(\frac{\phi_1(t)+\phi_3(t)}{2}-\phi_2(t)\Big)
\label{eq:signal}
\end{equation} 
Here $C(t)$ is the signal envelope related to the coherence of the source and $\phi_i(t)$ are the phases accumulated by the different paths. Relative to path 2, paths 1 and 3 accumulate phase from their kinetic energies and thus $S(t)$ oscillates at a frequency of $8\omega_{\rm rec}$, where $\omega_{\rm rec}=\hbar k^2/2m$ is the recoil frequency and $m$ is the mass of the atom. Importantly, effects from mirror vibrations on the optical standing wave phases cancel in this expression. Distinct from earlier realizations \cite{gupt02,jami14}, a dramatic enhancement of phase accumulation of $\frac{1}{2} n^2 \omega_{\rm rec} T$ is achieved in this work using multiple acceleration pulses. 

Our atom source consists of ytterbium ($^{174}$Yb) BECs of $N_{\rm at} = 150,000$ atoms prepared in a crossed-beam optical dipole trap operating at $532\,$nm. After condensate formation, we decompress the trap to a mean frequency of $\bar{\omega}=2\pi\times 63\,$Hz. To reduce the density and atomic interactions further, we allow $2\,$ms time-of-flight after trap turn-off before beginning the interferometry sequence.    
    
Our atom optics consist of diffraction beams near the ${^1}S_0 \rightarrow {^3}P_1 $ ($\lambda_g=556\,$nm$\,=\,2\pi/k$, $\Gamma_g\,=\,2\pi \times 182\,$kHz) intercombination transition and a readout beam near the ${^1}S_0 \rightarrow {^1}P_1 $ ($\lambda_b\,=\,399\,$nm, $\Gamma_b\,=\,2\pi \times 28\,$MHz) transition, both derived from our laser cooling sources. The two diffraction beams are detuned from the atomic resonance by $\Delta_g/\Gamma_g\,\simeq\,+3500$ and counter-propagate horizontally to form a standing wave. 
Each beam is derived from the first diffraction order of a 200 MHz acousto-optic modulator (AOM) driven by an Analog Devices AD9910 direct digital synthesizer. Each AOM output is passed through a polarization-maintaining single-mode fiber. 
The diffraction beams have waists of $1.8\,$mm, allowing lattice depths up to 50$\hbar \omega_{\rm rec}$ for the available power. We stabilized the diffraction beam intensities by feeding back to the AOMs, keeping fluctuations in the diffraction pulse peak lattice depth to $\leq 2\%$.

The splitting pulse has a width of 7$\,\mu$s, within the Kapitza-Dirac regime \cite{gupt01}. The mirror pulse is a second-order Bragg $\pi-$pulse with Gaussian 1/e full-width 54$\,\mu$s and peak lattice depth 14$\hbar\omega_{\rm rec}$. Each acceleration pulse is a third-order Bragg $\pi-$pulse delivering $6\hbar k$ of momentum, with Gaussian 1/e full-width 54$\,\mu$s and peak lattice depth 26.6$\hbar\omega_{\rm rec}$. We accelerate the outer paths sequentially as shown in Fig.\,\ref{fig:fig1}(a,b) with successive 
pulses separated by 130$\,\mu$s. Although this acceleration scheme breaks the symmetric form of the interferometer, the suppression of systematic effects from the symmetry of the three-path geometry \cite{gupt02,jami14} is largely retained if the time between acceleration pulses for paths 1 and 3 is short, as in our case. We use light at $\lambda_b$ for the readout beam which Bragg reflects at 44 degrees from the $\lambda_g/2$ period matter-wave grating to form the CI signal, eliminating noise associated with stray reflections when using Bragg back-scattering at $\lambda_g$ as in earlier work \cite{jami14}.

\begin{figure}[b]
\includegraphics[width=\linewidth]{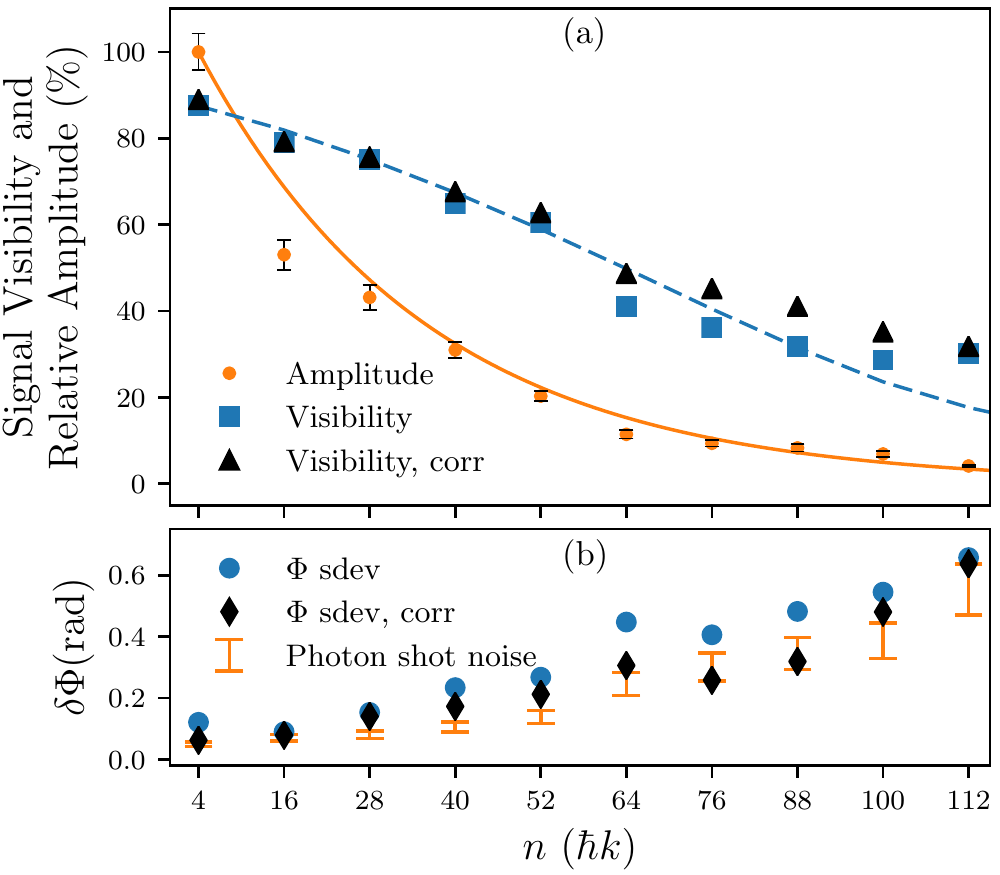}
\caption{(a) Normalized readout signal amplitude and visibility vs $n$. The amplitude is based on the Fourier component at $8\omega_{\rm rec}$, and visibility is extracted from sinusoidal fits to averaged data as shown in Fig.\,\ref{fig:fig1}. Signal visibility is calculated with (black triangles) and without (blue squares) the diffraction phase correction. The solid orange line is a fit to the observed amplitude and the dashed line is a model curve (see text). (b) CI phase standard deviation vs $n$, with (black diamonds) and without (blue circles) the diffraction phase correction. The orange bars show the expected limit from photon shot noise. The bar length, representing the estimation uncertainty, is dominated by a 30\% systematic uncertainty in the conversion of the PMT signal to photon number.}
\label{fig:fig2}
\end{figure}

Figure \ref{fig:fig1}(c-f) shows contrast readout signals for various values of $n$, each of which is an average of multiple experimental iterations (shots). The phase stability of the interferometer is apparent in the high visibility of these fringes even for the largest ($n=112$) momentum splitting used in this work. To our knowledge, this is the highest momentum splitting in any atom interferometer that produces stable, visible fringes. We attribute this capability to the vibration insensitivity of the CI and the suppression of diffraction phases discussed below. Note that these results are obtained without any active vibration isolation. To extract fringe visibility, we fit these signals with the expression $C(t_r) {\rm cos}^2(4\omega_{\rm rec}t_r+\Phi)+S_0$ using the currently-accepted value of $\omega_{\rm rec}$ and a Gaussian envelope $C(t_r)$ \cite{footgauss}. Here $t_r$ is the time from the start of the readout pulse and $S_0$ is a vertical offset. We quantify the visibility of our signal (Fig.\, \ref{fig:fig2}(a)) as [(Max-Min)/(Max+Min)]$\times 100\%$, where Max and Min are determined by our fitted values for $S_0$ and $C(t_r)$. The offset $S_0$ is due to the 7\% spontaneous scattering probability from the readout pulse, which is detuned by $\Delta_b/\Gamma_b=-50$. Note that this definition refers to the visibility of the fringe associated with light scattered from the atomic grating, and not to the visibility of the atomic grating itself, which is characterized by the amplitude of the scattered light presented in Figure \ref{fig:fig2}(a).

We analyze single shots by extracting the amplitude and phase ($2\Phi$) of the Fourier component at $8\omega_{\rm rec}$. Figure \ref{fig:fig2}(a) shows how the amplitude varies with $n$. Our data are well described by a simple model of imperfect acceleration pulses. The fraction of atoms remaining in path 1 or 3 that contribute to the final CI signal is $A^{\zeta}$, where $A$ is the efficiency per $\hbar k$ of our acceleration pulses, and $\zeta=\frac{4(n-4)}{2}$ is the total number of photon recoils from the acceleration pulses only, for path 1 or 3. A fit to the amplitude data using this model 
returns $A=0.9845(2)$, or 91\% per third-order Bragg pulse, consistent with a direct measurement of our $\pi$-pulse efficiency from absorption imaging of the atoms. This amplitude model, together with the $n$-dependent signal offset from spontaneous scattering, yields a visibility model that captures the main features of our data (dashed blue line).

\begin{figure}
  \includegraphics[width=\linewidth]{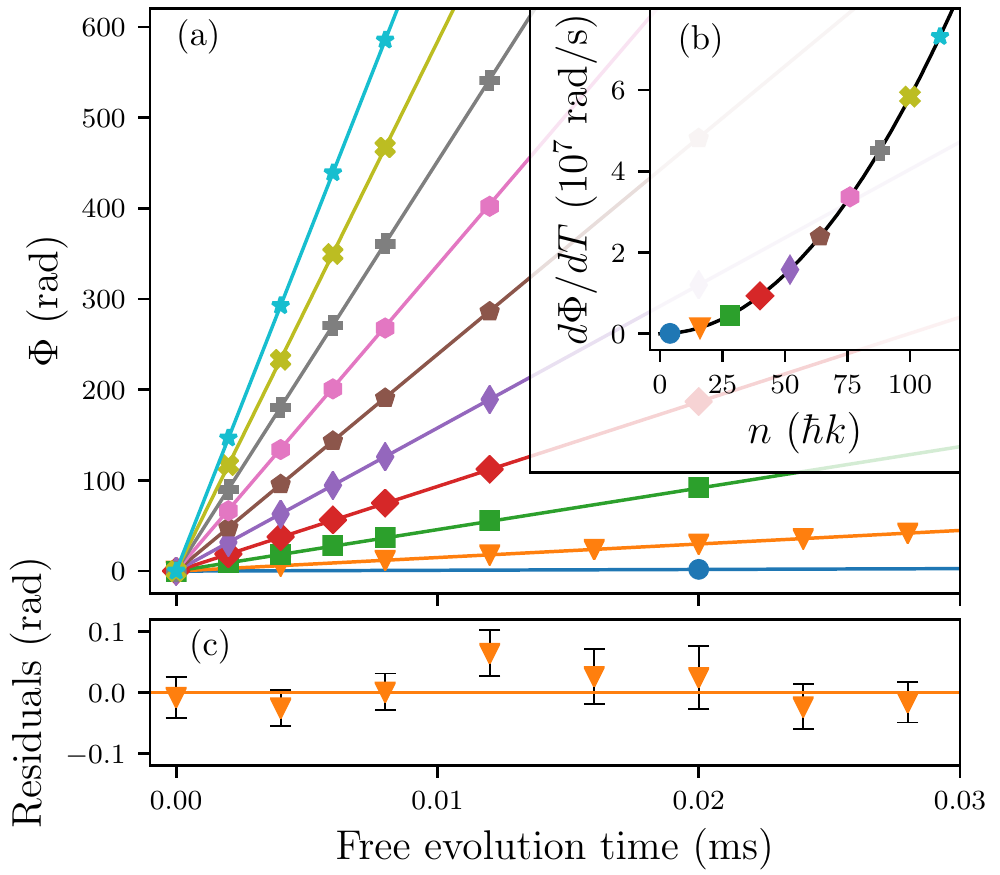}
  \caption{(a) CI phase $\Phi$ vs free evolution time $2T$ with linear fit, for various $n$. (b) Fit slopes $d\Phi/dT$ vs $n$, demonstrating the expected quadratic relationship $d\Phi/dT =\frac{1}{2} n^2\omega_\text{rec}$ (black curve). (c) Typical fit residuals ($n=16$). }
  \label{fig:fig3}
\end{figure}   

We characterize our interferometer's phase stability (important for precision measurements) as the standard deviation $\delta \Phi$ of extracted single-shot phases (Fig.\,\ref{fig:fig2}(b)). Our observations are close to the expectations from photon shot noise (orange bars), evaluated as $\sqrt{N_{\rm ph}}/({\sqrt{2}N_{\rm sig}})$ where $N_{\rm ph}$ is the average total number of detected photons and $N_{\rm sig}$ is the number contributing to the oscillatory part of the signal only \cite{lene97, footshot}. Since $N_{\rm ph}\!\ll\!N_{\rm at}$, the atom shot noise limit is far lower.  

For a free evolution time of $2T$, we define the signal phase at the start of the readout pulse to be the CI phase $\Phi(2T) = \frac{1}{2} n^2\omega_\text{rec}T + \Phi_{\rm offset}$. Here $\Phi_{\rm offset}$ contains a number of phase shifts that are common to interferometers of different $T$, as well as contributions from systematic effects. The evolution of $\Phi$ with $T$ is shown in Fig.\,\ref{fig:fig3} for various $n$. The fitted slopes are in good agreement with the expected $d\Phi/dT=\frac{1}{2} n^2\omega_\text{rec}$. We note that the interferometer phase at $n\!=\!112$ corresponds to the phase difference between two paths separated by $56\hbar k$ (Eqn.\ref{eq:signal}). 

The quadratic scaling of the CI phase with $n$ is a distinct benefit for precision measurements, however it comes at the cost of a systematic effect from diffraction phases. This effect stems from momentum-dependent phase shifts during Bragg diffraction and can be significant for different interferometer geometries \cite{buch03, jami14, este15, gies16}. A critical gauge of the viability of the CI scheme is the scaling of diffraction phase with momentum separation. An important element for our favorable $n$ scaling was the selection of acceleration pulse parameters that suppressed diffraction phase contributions to $\delta \Phi$, in addition to providing good atom optics efficiency.  

\begin{figure}[t]
  \includegraphics[width=\linewidth]{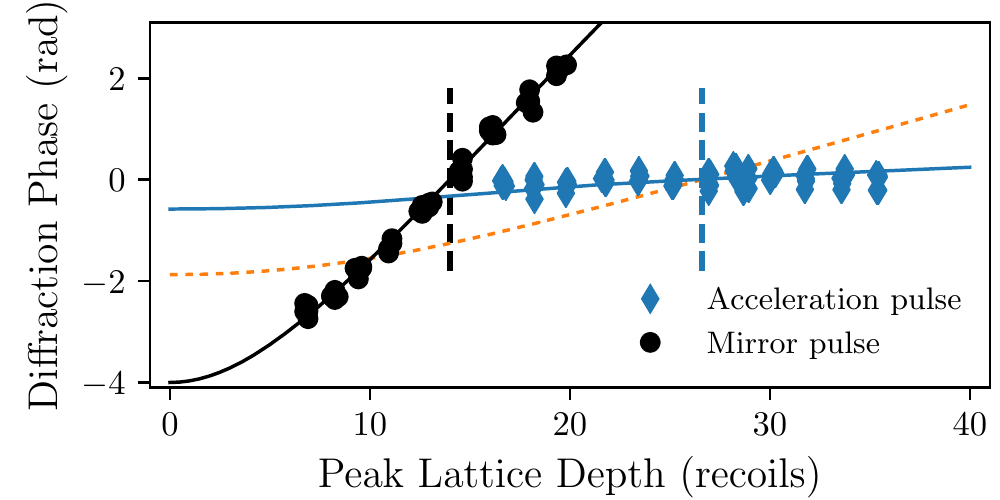}
  \caption{Diffraction phase shift vs peak lattice depth (pulse width held fixed) for the second-order Bragg mirror pulse (black circles) and the third-order acceleration pulse (blue diamonds) from $2\hbar k$ to $8\hbar k$. The dashed vertical lines indicate the peak lattice depths at which the $\pi-$pulse condition is met in each case. Overall phase offsets have been removed to zero the diffraction phases at the $\pi$-points. Black and blue solid lines are the predictions from the corresponding numerical model. The dotted orange line is the prediction of the model for the third-order Bragg pulse at large $n$ (see text).}
  \label{fig:fig4}
\end{figure}

The presence of an optical lattice modifies the atomic dispersion relations, leading to momentum-dependent (and therefore path-dependent) phase shifts which affect the CI signal according to Eqn.\ref{eq:signal}. We experimentally characterized the diffraction phase effect for our acceleration and mirror pulse parameters by varying the peak pulse intensity around the $\pi-$pulse condition ($\pi-$point) and observing the variation of the CI phase. As shown in Fig.\,\ref{fig:fig4}, our observations agree well with a numerical model of the diffraction processes which is equivalent to those described in \cite{jami14,este15}. The diffraction phase from the splitting pulse is negligible and the diffraction phases at the $\pi-$points for the Bragg mirror and acceleration pulses constitute a $T$-independent offset to the CI phase. Thus, in standard interferometer operation we need only consider the variations of the diffraction phase around the $\pi-$point from intensity fluctuations of our lattice. The black curve in Fig.\,\ref{fig:fig4} implies that 2\% intensity variations (our upper bound) in the mirror pulse contribute 70 mrad to the CI phase standard deviation. The first acceleration pulse contributes about $25\,$mrad (blue curve), while for larger $n$ the behavior converges to the dotted orange curve. For the largest $n=112$ interferometer there is less than $200\,$mrad diffraction phase fluctuations per shot, including the effects from all the pulses. Operation of the interferometer with the chosen parameters is crucial for its scalability. For instance, pulse widths four times longer lead to order of magnitude greater diffraction phases, which we verified both experimentally and theoretically. 

The CI phase fluctuations can be improved by applying a shot-by-shot diffraction phase correction based on the correlation between the CI phase and the recorded diffraction pulse amplitudes. The correction is significant, reducing $\delta \Phi$ to $<320\,$mrad for $n$ up to 88 (Fig.\,\ref{fig:fig2}(b)) and bringing our observations into closer agreement with the photon shot noise limit. We also observe a small improvement in the visibilities of the corrected averaged data (Fig.\,\ref{fig:fig2}(a)).

We have considered the effects from atomic interactions on our results. For our Thomas-Fermi condensate source \cite{cast96,jami11} the estimated phase fluctuations arising from the $<\!3\%$ fluctuations in initial splitting asymmetry are far less than those observed.
    
We now consider application of the large $n$ CI technique to a photon recoil and $\alpha$ measurement. The precision in $\omega_{\rm rec}$ can be written as: 
\begin{equation}
\frac{\delta \omega_{\rm rec}}{\omega_{\rm rec}}=\frac{\delta \Phi}{\Phi}= \frac{\delta \Phi}{\frac{1}{2} n^2\omega_{\text{rec}}\Delta T\sqrt{M}} 
\label{eq:precision}
\end{equation}
where $2\Delta T$ is the range of free evolution times over which the slope of $\Phi(2T)$ is measured and $M$ is the number of experimental shots. In our current CI setup, the free evolution time is constrained by the atoms falling out of the horizontally oriented diffraction beams, and $\frac{1}{2}n^2\omega_\text{rec}T$ is optimized to $2.1 \times 10^5$ radians for $n=76$, and $T=3\,$ms. This represents an improvement of two orders of magnitude in total interferometer phase compared to our earlier CI realization \cite{jami14}. For these parameters, the maximum separation of interfering states is 1.5mm. The observed $\delta\Phi$=250\,mrad at $n=76$ (Fig.\,\ref{fig:fig2}(b)) then gives a precision of $8.7\times10^{-8}$ in $\omega_\text{rec}$ in 200 shots.

The interferometer cycle time is dominated by BEC production. While this is $10\,$s for this work, we have demonstrated Yb BEC cycle times as low as $1.6\,$s in our group \cite{roy16}. Using $3\,$s as a reasonable benchmark for longterm measurements, the above numbers scale to $1.1 \times 10^{-8}$ in $\omega_{\rm rec}$ in 10 hrs of integration time. We are initiating a new CI configuration with vertically oriented diffraction beams where the limitation on free evolution time is lifted and $2T=210\,$ms (keeping $n=76$) is possible in a $7\,$cm vertical region. We have also demonstrated delta-kick cooling \cite{munt13, kova15} in our experiment which will help preserve the interferometer signal quality for large $T$. The above scaling then indicates a precision of $3.2\times10^{-10}$ in $\omega_{\rm rec}$ in 10 hrs \cite{footsys}. The corresponding precision in $\alpha$, which can be determined by combining $\omega_{\rm rec}$ and measurements of other fundamental constants \cite{weis93} is a factor of two better. Together with potential improvements in $n$ and $\delta \Phi$ from better interferometer pulse control, this approach holds promise for a $10^{-10}$ level measurement of $\alpha$ and test of QED \cite{hann08,bouc11,park18,aoya12}.           

Even though the CI signal is insensitive to acceleration, it is sensitive to its first derivative \cite{jami14thesis}, and thus to gravity gradients. Our techniques 
for large $n$ interferometers should therefore also positively impact other applications of atom interferometry, including gravity gradiometry \cite{mcgu02} and measurement of the Newtonian gravitational constant \cite{fixl07,rosi14}.   
  
In summary, we have developed a 
high-visibility phase-stable atom interferometer with momentum splitting up to $112\hbar k$, exceeding the momentum separation achieved in earlier phase-stable free-space interferometers. The robust scalability arises from the inherent vibration insensitivity of the interferometer geometry as well as diffraction phase control. 
We demonstrated a quadratic growth of interferometer phase with momentum splitting and favorable scaling of the performance towards a precision measurement of $\alpha$. Finally, our results also represent an important advance in the use of alkaline-earth-like atoms for precision atom interferometry, where their ground-state magnetic field insensitivity and the presence of narrow intercombination transitions can be exploited \cite{rieh91,jami14,tara14,hupo17,agui17,grah13,hart15,norc17}.

\begin{acknowledgments}
We thank Brendan Saxberg and Ryan Weh for important technical contributions. E.S.C. acknowledges support from the NSF-REU program at the University of Washington. This work was supported by NSF Grants No. PHY-1404075 and PHY-1707575.
\end{acknowledgments}


\begin{thebibliography}{45}
\expandafter\ifx\csname natexlab\endcsname\relax\def\natexlab#1{#1}\fi
\expandafter\ifx\csname bibnamefont\endcsname\relax
  \def\bibnamefont#1{#1}\fi
\expandafter\ifx\csname bibfnamefont\endcsname\relax
  \def\bibfnamefont#1{#1}\fi
\expandafter\ifx\csname citenamefont\endcsname\relax
  \def\citenamefont#1{#1}\fi
\expandafter\ifx\csname url\endcsname\relax
  \def\url#1{\texttt{#1}}\fi
\expandafter\ifx\csname urlprefix\endcsname\relax\def\urlprefix{URL }\fi
\providecommand{\bibinfo}[2]{#2}
\providecommand{\eprint}[2][]{\url{#2}}

\bibitem[{\citenamefont{Cronin et~al.}(2009)\citenamefont{Cronin, Schmiedmayer,
  and Pritchard}}]{cron09}
\bibinfo{author}{\bibfnamefont{A.}~\bibnamefont{Cronin}},
  \bibinfo{author}{\bibfnamefont{J.}~\bibnamefont{Schmiedmayer}},
  \bibnamefont{and}
  \bibinfo{author}{\bibfnamefont{D.}~\bibnamefont{Pritchard}},
  \bibinfo{journal}{Rev. Mod. Phys.} \textbf{\bibinfo{volume}{81}},
  \bibinfo{pages}{1051} (\bibinfo{year}{2009}).

\bibitem[{\citenamefont{Peters et~al.}(2001)\citenamefont{Peters, Chung, and
  Chu}}]{pete01}
\bibinfo{author}{\bibfnamefont{A.}~\bibnamefont{Peters}},
  \bibinfo{author}{\bibfnamefont{K.}~\bibnamefont{Chung}}, \bibnamefont{and}
  \bibinfo{author}{\bibfnamefont{S.}~\bibnamefont{Chu}},
  \bibinfo{journal}{Metrologia} \textbf{\bibinfo{volume}{38}},
  \bibinfo{pages}{25} (\bibinfo{year}{2001}).

\bibitem[{\citenamefont{McGuirk et~al.}(2002)\citenamefont{McGuirk, Foster,
  Fixler, Snadden, and Kasevich}}]{mcgu02}
\bibinfo{author}{\bibfnamefont{J.}~\bibnamefont{McGuirk}},
  \bibinfo{author}{\bibfnamefont{G.}~\bibnamefont{Foster}},
  \bibinfo{author}{\bibfnamefont{J.}~\bibnamefont{Fixler}},
  \bibinfo{author}{\bibfnamefont{M.}~\bibnamefont{Snadden}}, \bibnamefont{and}
  \bibinfo{author}{\bibfnamefont{M.}~\bibnamefont{Kasevich}},
  \bibinfo{journal}{Phys. Rev. A} \textbf{\bibinfo{volume}{65}},
  \bibinfo{pages}{033608} (\bibinfo{year}{2002}).

\bibitem[{\citenamefont{Durfee et~al.}(2006)\citenamefont{Durfee, Shaham, and
  Kasevich}}]{durf06}
\bibinfo{author}{\bibfnamefont{D.}~\bibnamefont{Durfee}},
  \bibinfo{author}{\bibfnamefont{Y.}~\bibnamefont{Shaham}}, \bibnamefont{and}
  \bibinfo{author}{\bibfnamefont{M.}~\bibnamefont{Kasevich}},
  \bibinfo{journal}{Phys. Rev. Lett.} \textbf{\bibinfo{volume}{97}},
  \bibinfo{pages}{240801} (\bibinfo{year}{2006}).

\bibitem[{\citenamefont{Dutta et~al.}(2016)\citenamefont{Dutta, Savoie, Fang,
  Venon, GarridoAlzar, Geiger, and Landragin}}]{dutt16}
\bibinfo{author}{\bibfnamefont{I.}~\bibnamefont{Dutta}},
  \bibinfo{author}{\bibfnamefont{D.}~\bibnamefont{Savoie}},
  \bibinfo{author}{\bibfnamefont{B.}~\bibnamefont{Fang}},
  \bibinfo{author}{\bibfnamefont{B.}~\bibnamefont{Venon}},
  \bibinfo{author}{\bibfnamefont{C.}~\bibnamefont{GarridoAlzar}},
  \bibinfo{author}{\bibfnamefont{R.}~\bibnamefont{Geiger}}, \bibnamefont{and}
  \bibinfo{author}{\bibfnamefont{A.}~\bibnamefont{Landragin}},
  \bibinfo{journal}{Phys. Rev. Lett.} \textbf{\bibinfo{volume}{116}},
  \bibinfo{pages}{183003} (\bibinfo{year}{2016}).

\bibitem[{\citenamefont{Geiger et~al.}(2011)\citenamefont{Geiger, Menoret,
  Stern, Zahzam, Cheinet, Battelier, Villing, Moron, Lours, Bidel
  et~al.}}]{geig11}
\bibinfo{author}{\bibfnamefont{R.}~\bibnamefont{Geiger}},
  \bibinfo{author}{\bibfnamefont{V.}~\bibnamefont{Menoret}},
  \bibinfo{author}{\bibfnamefont{G.}~\bibnamefont{Stern}},
  \bibinfo{author}{\bibfnamefont{N.}~\bibnamefont{Zahzam}},
  \bibinfo{author}{\bibfnamefont{P.}~\bibnamefont{Cheinet}},
  \bibinfo{author}{\bibfnamefont{B.}~\bibnamefont{Battelier}},
  \bibinfo{author}{\bibfnamefont{A.}~\bibnamefont{Villing}},
  \bibinfo{author}{\bibfnamefont{F.}~\bibnamefont{Moron}},
  \bibinfo{author}{\bibfnamefont{M.}~\bibnamefont{Lours}},
  \bibinfo{author}{\bibfnamefont{Y.}~\bibnamefont{Bidel}},
  \bibnamefont{et~al.}, \bibinfo{journal}{Nat. Commun.}
  \textbf{\bibinfo{volume}{2}}, \bibinfo{pages}{474} (\bibinfo{year}{2011}).

\bibitem[{\citenamefont{Fray et~al.}(2004)\citenamefont{Fray, Diez, Hänsch,
  and Weitz}}]{fray04}
\bibinfo{author}{\bibfnamefont{S.}~\bibnamefont{Fray}},
  \bibinfo{author}{\bibfnamefont{C.}~\bibnamefont{Diez}},
  \bibinfo{author}{\bibfnamefont{T.}~\bibnamefont{Hänsch}}, \bibnamefont{and}
  \bibinfo{author}{\bibfnamefont{M.}~\bibnamefont{Weitz}},
  \bibinfo{journal}{Phys. Rev. Lett.} \textbf{\bibinfo{volume}{93}},
  \bibinfo{pages}{240404} (\bibinfo{year}{2004}).

\bibitem[{\citenamefont{Schlippert et~al.}(2014)\citenamefont{Schlippert,
  Hartwig, Albers, Richardson, Schubert, Roura, Schleich, Ertmer, and
  Rasel}}]{schl14}
\bibinfo{author}{\bibfnamefont{D.}~\bibnamefont{Schlippert}},
  \bibinfo{author}{\bibfnamefont{J.}~\bibnamefont{Hartwig}},
  \bibinfo{author}{\bibfnamefont{H.}~\bibnamefont{Albers}},
  \bibinfo{author}{\bibfnamefont{L.}~\bibnamefont{Richardson}},
  \bibinfo{author}{\bibfnamefont{C.}~\bibnamefont{Schubert}},
  \bibinfo{author}{\bibfnamefont{A.}~\bibnamefont{Roura}},
  \bibinfo{author}{\bibfnamefont{W.}~\bibnamefont{Schleich}},
  \bibinfo{author}{\bibfnamefont{W.}~\bibnamefont{Ertmer}}, \bibnamefont{and}
  \bibinfo{author}{\bibfnamefont{E.}~\bibnamefont{Rasel}},
  \bibinfo{journal}{Phys. Rev. Lett.} \textbf{\bibinfo{volume}{112}},
  \bibinfo{pages}{203002} (\bibinfo{year}{2014}).

\bibitem[{\citenamefont{Bouchendira et~al.}(2011)\citenamefont{Bouchendira,
  Clade, Guellati-Khelifa, Nez, and Biraben}}]{bouc11}
\bibinfo{author}{\bibfnamefont{R.}~\bibnamefont{Bouchendira}},
  \bibinfo{author}{\bibfnamefont{P.}~\bibnamefont{Clade}},
  \bibinfo{author}{\bibfnamefont{S.}~\bibnamefont{Guellati-Khelifa}},
  \bibinfo{author}{\bibfnamefont{F.}~\bibnamefont{Nez}}, \bibnamefont{and}
  \bibinfo{author}{\bibfnamefont{F.}~\bibnamefont{Biraben}},
  \bibinfo{journal}{Phys. Rev. Lett.} \textbf{\bibinfo{volume}{106}},
  \bibinfo{pages}{080801} (\bibinfo{year}{2011}).
  
\bibitem[{\citenamefont{Parker et~al.}(2018)\citenamefont{Parker, Yu,
  Zhong, Estey, and Kasevich}}]{park18}
\bibinfo{author}{\bibfnamefont{R.}~\bibnamefont{Parker}},
  \bibinfo{author}{\bibfnamefont{C.}~\bibnamefont{Yu}},
  \bibinfo{author}{\bibfnamefont{W.}~\bibnamefont{Zhong}}, 
  \bibinfo{author}{\bibfnamefont{B.}~\bibnamefont{Estey}}, \bibnamefont{and}
  \bibinfo{author}{\bibfnamefont{H.}~\bibnamefont{Muller}},
  \bibinfo{journal}{Science} \textbf{\bibinfo{volume}{360}},
  \bibinfo{pages}{191} (\bibinfo{year}{2018}).

\bibitem[{\citenamefont{Muller et~al.}(2009)\citenamefont{Muller, Chiow,
  Herrmann, and Chu}}]{mull09}
\bibinfo{author}{\bibfnamefont{H.}~\bibnamefont{Muller}},
  \bibinfo{author}{\bibfnamefont{S.}~\bibnamefont{Chiow}},
  \bibinfo{author}{\bibfnamefont{S.}~\bibnamefont{Herrmann}}, \bibnamefont{and}
  \bibinfo{author}{\bibfnamefont{S.}~\bibnamefont{Chu}},
  \bibinfo{journal}{Phys. Rev. Lett.} \textbf{\bibinfo{volume}{102}},
  \bibinfo{pages}{240403} (\bibinfo{year}{2009}).

\bibitem[{\citenamefont{Chiow et~al.}(2011)\citenamefont{Chiow, Kovachy, Chien,
  and Kasevich}}]{chio11}
\bibinfo{author}{\bibfnamefont{S.}~\bibnamefont{Chiow}},
  \bibinfo{author}{\bibfnamefont{T.}~\bibnamefont{Kovachy}},
  \bibinfo{author}{\bibfnamefont{H.}~\bibnamefont{Chien}}, \bibnamefont{and}
  \bibinfo{author}{\bibfnamefont{M.~A.} \bibnamefont{Kasevich}},
  \bibinfo{journal}{Phys. Rev. Lett.} \textbf{\bibinfo{volume}{107}},
  \bibinfo{pages}{130403} (\bibinfo{year}{2011}).

\bibitem[{foo(2017{\natexlab{a}})}]{footkas}
\bibinfo{journal}{We use the term phase stability to mean that the shot to shot
  variation in phase is less than 2$\pi$, i.e. reproducible and 
  usable for a direct interferometric measurement}.

\bibitem[{\citenamefont{Chiow et~al.}(2009)\citenamefont{Chiow, Herrmann, Chu,
  and Muller}}]{chio09}
\bibinfo{author}{\bibfnamefont{S.}~\bibnamefont{Chiow}},
  \bibinfo{author}{\bibfnamefont{S.}~\bibnamefont{Herrmann}},
  \bibinfo{author}{\bibfnamefont{S.}~\bibnamefont{Chu}}, \bibnamefont{and}
  \bibinfo{author}{\bibfnamefont{H.}~\bibnamefont{Muller}},
  \bibinfo{journal}{Phys. Rev. Lett.} \textbf{\bibinfo{volume}{103}},
  \bibinfo{pages}{050402} (\bibinfo{year}{2009}).

\bibitem[{\citenamefont{Muller et~al.}(2008)\citenamefont{Muller, Chiow, Long,
  Herrmann, and Chu}}]{mull08}
\bibinfo{author}{\bibfnamefont{H.}~\bibnamefont{Muller}},
  \bibinfo{author}{\bibfnamefont{S.}~\bibnamefont{Chiow}},
  \bibinfo{author}{\bibfnamefont{Q.}~\bibnamefont{Long}},
  \bibinfo{author}{\bibfnamefont{S.}~\bibnamefont{Herrmann}}, \bibnamefont{and}
  \bibinfo{author}{\bibfnamefont{S.}~\bibnamefont{Chu}},
  \bibinfo{journal}{Phys. Rev. Lett.} \textbf{\bibinfo{volume}{100}},
  \bibinfo{pages}{180405} (\bibinfo{year}{2008}).

\bibitem[{\citenamefont{Asenbaum et~al.}(2017)\citenamefont{Asenbaum,
  Overstreet, Kovachy, Brown, Hogan, and Kasevich}}]{asen17}
\bibinfo{author}{\bibfnamefont{P.}~\bibnamefont{Asenbaum}},
  \bibinfo{author}{\bibfnamefont{C.}~\bibnamefont{Overstreet}},
  \bibinfo{author}{\bibfnamefont{T.}~\bibnamefont{Kovachy}},
  \bibinfo{author}{\bibfnamefont{D.}~\bibnamefont{Brown}},
  \bibinfo{author}{\bibfnamefont{J.}~\bibnamefont{Hogan}}, \bibnamefont{and}
  \bibinfo{author}{\bibfnamefont{M.}~\bibnamefont{Kasevich}},
  \bibinfo{journal}{Phys. Rev. Lett.} \textbf{\bibinfo{volume}{118}},
  \bibinfo{pages}{183602} (\bibinfo{year}{2017}).

\bibitem[{\citenamefont{McDonald et~al.}(2013)\citenamefont{McDonald, Kuhn,
  Bennetts, Debs, Hardman, Johnsson, Close, and Robins}}]{mcdo13}
\bibinfo{author}{\bibfnamefont{G.}~\bibnamefont{McDonald}},
  \bibinfo{author}{\bibfnamefont{C.}~\bibnamefont{Kuhn}},
  \bibinfo{author}{\bibfnamefont{S.}~\bibnamefont{Bennetts}},
  \bibinfo{author}{\bibfnamefont{J.}~\bibnamefont{Debs}},
  \bibinfo{author}{\bibfnamefont{K.}~\bibnamefont{Hardman}},
  \bibinfo{author}{\bibfnamefont{M.}~\bibnamefont{Johnsson}},
  \bibinfo{author}{\bibfnamefont{J.}~\bibnamefont{Close}}, \bibnamefont{and}
  \bibinfo{author}{\bibfnamefont{N.}~\bibnamefont{Robins}},
  \bibinfo{journal}{Phys. Rev. A} \textbf{\bibinfo{volume}{88}},
  \bibinfo{pages}{053620} (\bibinfo{year}{2013}).

\bibitem[{\citenamefont{Gupta et~al.}(2002)\citenamefont{Gupta, Dieckmann,
  Hadzibabic, and Pritchard}}]{gupt02}
\bibinfo{author}{\bibfnamefont{S.}~\bibnamefont{Gupta}},
  \bibinfo{author}{\bibfnamefont{K.}~\bibnamefont{Dieckmann}},
  \bibinfo{author}{\bibfnamefont{Z.}~\bibnamefont{Hadzibabic}},
  \bibnamefont{and} \bibinfo{author}{\bibfnamefont{D.~E.}
  \bibnamefont{Pritchard}}, \bibinfo{journal}{Phys. Rev. Lett.}
  \textbf{\bibinfo{volume}{89}}, \bibinfo{pages}{140401}
  (\bibinfo{year}{2002}).

\bibitem[{\citenamefont{Jamison et~al.}(2014)\citenamefont{Jamison,
  Plotkin-Swing, and Gupta}}]{jami14}
\bibinfo{author}{\bibfnamefont{A.~O.} \bibnamefont{Jamison}},
  \bibinfo{author}{\bibfnamefont{B.}~\bibnamefont{Plotkin-Swing}},
  \bibnamefont{and} \bibinfo{author}{\bibfnamefont{S.}~\bibnamefont{Gupta}},
  \bibinfo{journal}{Phys. Rev. A} \textbf{\bibinfo{volume}{90}},
  \bibinfo{pages}{063606} (\bibinfo{year}{2014}).

\bibitem[{foo(2017{\natexlab{b}})}]{foot4hk}
\bibinfo{journal}{An additional matter-wave grating formed by the interference
  of only paths 1 and 3 has period $\pi/2k$ and therefore does not reflect our
  probe beam.} 

\bibitem[{\citenamefont{Gupta et~al.}(2001)\citenamefont{Gupta, Leanhardt,
  Cronin, and Pritchard}}]{gupt01}
\bibinfo{author}{\bibfnamefont{S.}~\bibnamefont{Gupta}},
  \bibinfo{author}{\bibfnamefont{A.~E.} \bibnamefont{Leanhardt}},
  \bibinfo{author}{\bibfnamefont{A.~D.} \bibnamefont{Cronin}},
  \bibnamefont{and} \bibinfo{author}{\bibfnamefont{D.~E.}
  \bibnamefont{Pritchard}}, \bibinfo{journal}{Cr. Acad. Sci. IV-Phys}
  \textbf{\bibinfo{volume}{2}}, \bibinfo{pages}{479} (\bibinfo{year}{2001}).

\bibitem[{foo(2017{\natexlab{c}})}]{footgauss}
\bibinfo{journal}{The width of the readout signal is determined by the
  coherence time $1/k\Delta v$ where $\Delta v$ is the velocity spread of the
  atom source \cite{gupt02}. The visibility in the central part of the signal
  is insensitive to the choice of envelope function.} 


\bibitem[{\citenamefont{Lenef et~al.}(1997)\citenamefont{Lenef, Hammond, Smith, Chapman, Rubenstein, and Pritchard}}]{lene97}
\bibinfo{author}{\bibfnamefont{A.}~\bibnamefont{Lenef}},
  \bibinfo{author}{\bibfnamefont{T.}~\bibnamefont{Hammond}},
  \bibinfo{author}{\bibfnamefont{E.}~\bibnamefont{Smith}},
  \bibinfo{author}{\bibfnamefont{M.}~\bibnamefont{Chapman}},
  \bibinfo{author}{\bibfnamefont{R.}~\bibnamefont{Rubenstein}}, \bibnamefont{and}
  \bibinfo{author}{\bibfnamefont{D.}~\bibnamefont{Pritchard}},
  \bibinfo{journal}{Phys. Rev. Lett} \textbf{\bibinfo{volume}{78}},
  \bibinfo{pages}{760} (\bibinfo{year}{1997}).

\bibitem[{foo(2017{\natexlab{e}})}]{footshot}
\bibinfo{journal}{Our expression is a factor of 2 lower than the one in \cite{lene97} reflecting the phase of the ${\rm cos}^2(4\omega_{\rm rec}t)$ function rather than that of ${\rm cos}(8\omega_{\rm rec}t)$.}

\bibitem[{\citenamefont{Buchner et~al.}(2003)\citenamefont{Buchner, Delhuille,
  Miffre, Robilliard, Vigue, and Champenois}}]{buch03}
\bibinfo{author}{\bibfnamefont{M.}~\bibnamefont{Buchner}},
  \bibinfo{author}{\bibfnamefont{R.}~\bibnamefont{Delhuille}},
  \bibinfo{author}{\bibfnamefont{A.}~\bibnamefont{Miffre}},
  \bibinfo{author}{\bibfnamefont{C.}~\bibnamefont{Robilliard}},
  \bibinfo{author}{\bibfnamefont{J.}~\bibnamefont{Vigue}}, \bibnamefont{and}
  \bibinfo{author}{\bibfnamefont{C.}~\bibnamefont{Champenois}},
  \bibinfo{journal}{Phys. Rev. A} \textbf{\bibinfo{volume}{68}},
  \bibinfo{pages}{013607} (\bibinfo{year}{2003}).

\bibitem[{\citenamefont{Estey et~al.}(2015)\citenamefont{Estey, Yu, Muller,
  Kuan, and Lan}}]{este15}
\bibinfo{author}{\bibfnamefont{B.}~\bibnamefont{Estey}},
  \bibinfo{author}{\bibfnamefont{C.}~\bibnamefont{Yu}},
  \bibinfo{author}{\bibfnamefont{H.}~\bibnamefont{Muller}},
  \bibinfo{author}{\bibfnamefont{P.}~\bibnamefont{Kuan}}, \bibnamefont{and}
  \bibinfo{author}{\bibfnamefont{S.}~\bibnamefont{Lan}},
  \bibinfo{journal}{Phys. Rev. Lett.} \textbf{\bibinfo{volume}{115}},
  \bibinfo{pages}{083002} (\bibinfo{year}{2015}).

\bibitem[{\citenamefont{Giese et~al.}(2016)\citenamefont{Giese, Friedrich,
  Abend, Rasel, and Schleich}}]{gies16}
\bibinfo{author}{\bibfnamefont{E.}~\bibnamefont{Giese}},
  \bibinfo{author}{\bibfnamefont{A.}~\bibnamefont{Friedrich}},
  \bibinfo{author}{\bibfnamefont{S.}~\bibnamefont{Abend}},
  \bibinfo{author}{\bibfnamefont{E.}~\bibnamefont{Rasel}}, \bibnamefont{and}
  \bibinfo{author}{\bibfnamefont{W.}~\bibnamefont{Schleich}},
  \bibinfo{journal}{Phys. Rev. A.} \textbf{\bibinfo{volume}{94}},
  \bibinfo{pages}{063619} (\bibinfo{year}{2016}).

\bibitem[{\citenamefont{Castin and Dum}(1996)}]{cast96}
\bibinfo{author}{\bibfnamefont{Y.}~\bibnamefont{Castin}} \bibnamefont{and}
  \bibinfo{author}{\bibfnamefont{R.}~\bibnamefont{Dum}},
  \bibinfo{journal}{Phys. Rev. Lett.} \textbf{\bibinfo{volume}{77}},
  \bibinfo{pages}{5315} (\bibinfo{year}{1996}).

\bibitem[{\citenamefont{Jamison et~al.}(2011)\citenamefont{Jamison, Kutz, and
  Gupta}}]{jami11}
\bibinfo{author}{\bibfnamefont{A.~O.} \bibnamefont{Jamison}},
  \bibinfo{author}{\bibfnamefont{J.~N.} \bibnamefont{Kutz}}, \bibnamefont{and}
  \bibinfo{author}{\bibfnamefont{S.}~\bibnamefont{Gupta}},
  \bibinfo{journal}{Phys. Rev. A.} \textbf{\bibinfo{volume}{84}},
  \bibinfo{pages}{043643} (\bibinfo{year}{2011}).

\bibitem[{\citenamefont{Roy et~al.}(2016)\citenamefont{Roy, Green, Bowler, and
  Gupta}}]{roy16}
\bibinfo{author}{\bibfnamefont{R.}~\bibnamefont{Roy}},
  \bibinfo{author}{\bibfnamefont{A.}~\bibnamefont{Green}},
  \bibinfo{author}{\bibfnamefont{R.}~\bibnamefont{Bowler}}, \bibnamefont{and}
  \bibinfo{author}{\bibfnamefont{S.}~\bibnamefont{Gupta}},
  \bibinfo{journal}{Phys. Rev. A.} \textbf{\bibinfo{volume}{93}},
  \bibinfo{pages}{043403} (\bibinfo{year}{2016}).

\bibitem[{\citenamefont{Muntinga et. al.}(2013)}]{munt13}
\bibinfo{author}{\bibfnamefont{H.}~\bibnamefont{Muntinga}}
  \bibinfo{author}{\bibnamefont{et. al.}}, \bibinfo{journal}{Phys. Rev. Lett.}
  \textbf{\bibinfo{volume}{110}}, \bibinfo{pages}{093602}
  (\bibinfo{year}{2013}).

\bibitem[{\citenamefont{Kovachy et~al.}(2015)\citenamefont{Kovachy, Hogan,
  Sugarbaker, Dickerson, Donnelly, Overstreet, and Kasevich}}]{kova15}
\bibinfo{author}{\bibfnamefont{T.}~\bibnamefont{Kovachy}},
  \bibinfo{author}{\bibfnamefont{J.}~\bibnamefont{Hogan}},
  \bibinfo{author}{\bibfnamefont{A.}~\bibnamefont{Sugarbaker}},
  \bibinfo{author}{\bibfnamefont{S.}~\bibnamefont{Dickerson}},
  \bibinfo{author}{\bibfnamefont{C.}~\bibnamefont{Donnelly}},
  \bibinfo{author}{\bibfnamefont{C.}~\bibnamefont{Overstreet}},
  \bibnamefont{and} \bibinfo{author}{\bibfnamefont{M.}~\bibnamefont{Kasevich}},
  \bibinfo{journal}{Phys. Rev. Lett.} \textbf{\bibinfo{volume}{114}},
  \bibinfo{pages}{143004} (\bibinfo{year}{2015}).

\bibitem[{foo(2017{\natexlab{f}})}]{footsys}
\bibinfo{journal}{An analysis of relevant systematic effects can be found in \cite{jami14thesis}.}

\bibitem[{\citenamefont{Weiss et~al.}(1993)\citenamefont{Weiss, Young, and
  Chu}}]{weis93}
\bibinfo{author}{\bibfnamefont{D.~S.} \bibnamefont{Weiss}},
  \bibinfo{author}{\bibfnamefont{B.~C.} \bibnamefont{Young}}, \bibnamefont{and}
  \bibinfo{author}{\bibfnamefont{S.}~\bibnamefont{Chu}},
  \bibinfo{journal}{Phys. Rev. Lett.} \textbf{\bibinfo{volume}{70}},
  \bibinfo{pages}{2706} (\bibinfo{year}{1993}).

\bibitem[{\citenamefont{Hanneke et~al.}(2008)\citenamefont{Hanneke, Fogwell,
  and Gabrielse}}]{hann08}
\bibinfo{author}{\bibfnamefont{D.}~\bibnamefont{Hanneke}},
  \bibinfo{author}{\bibfnamefont{S.}~\bibnamefont{Fogwell}}, \bibnamefont{and}
  \bibinfo{author}{\bibfnamefont{G.}~\bibnamefont{Gabrielse}},
  \bibinfo{journal}{Phys. Rev. Lett.} \textbf{\bibinfo{volume}{100}},
  \bibinfo{pages}{120801} (\bibinfo{year}{2008}).

\bibitem[{\citenamefont{Aoyama et~al.}(2012)\citenamefont{Aoyama, Hayakawa,
  Kinoshita, and Nio}}]{aoya12}
\bibinfo{author}{\bibfnamefont{T.}~\bibnamefont{Aoyama}},
  \bibinfo{author}{\bibfnamefont{M.}~\bibnamefont{Hayakawa}},
  \bibinfo{author}{\bibfnamefont{T.}~\bibnamefont{Kinoshita}},
  \bibnamefont{and} \bibinfo{author}{\bibfnamefont{M.}~\bibnamefont{Nio}},
  \bibinfo{journal}{Phys. Rev. Lett.} \textbf{\bibinfo{volume}{109}},
  \bibinfo{pages}{111807} (\bibinfo{year}{2012}).

\bibitem[{\citenamefont{Jamison}(2014)}]{jami14thesis}
\bibinfo{author}{\bibfnamefont{A.~O.} \bibnamefont{Jamison}},
  \bibinfo{journal}{University of Washington PhD Thesis}
  (\bibinfo{year}{2014}).

\bibitem[{\citenamefont{Fixler et~al.}(2007)\citenamefont{Fixler, Foster,
  McGuirk, and Kasevich}}]{fixl07}
\bibinfo{author}{\bibfnamefont{J.}~\bibnamefont{Fixler}},
  \bibinfo{author}{\bibfnamefont{G.}~\bibnamefont{Foster}},
  \bibinfo{author}{\bibfnamefont{J.}~\bibnamefont{McGuirk}}, \bibnamefont{and}
  \bibinfo{author}{\bibfnamefont{M.}~\bibnamefont{Kasevich}},
  \bibinfo{journal}{Science} \textbf{\bibinfo{volume}{315}},
  \bibinfo{pages}{74} (\bibinfo{year}{2007}).

\bibitem[{\citenamefont{Rosi et~al.}(2014)\citenamefont{Rosi, Sorrentino,
  Cacciapuoti, Prevedelli, and Tino}}]{rosi14}
\bibinfo{author}{\bibfnamefont{G.}~\bibnamefont{Rosi}},
  \bibinfo{author}{\bibfnamefont{F.}~\bibnamefont{Sorrentino}},
  \bibinfo{author}{\bibfnamefont{L.}~\bibnamefont{Cacciapuoti}},
  \bibinfo{author}{\bibfnamefont{M.}~\bibnamefont{Prevedelli}},
  \bibnamefont{and} \bibinfo{author}{\bibfnamefont{G.}~\bibnamefont{Tino}},
  \bibinfo{journal}{Nature} \textbf{\bibinfo{volume}{510}},
  \bibinfo{pages}{518} (\bibinfo{year}{2014}).

\bibitem[{\citenamefont{Riehle et~al.}(1991)\citenamefont{Riehle, Kisters,
  Witte, Helmcke, and Borde}}]{rieh91}
\bibinfo{author}{\bibfnamefont{F.}~\bibnamefont{Riehle}},
  \bibinfo{author}{\bibfnamefont{T.}~\bibnamefont{Kisters}},
  \bibinfo{author}{\bibfnamefont{A.}~\bibnamefont{Witte}},
  \bibinfo{author}{\bibfnamefont{J.}~\bibnamefont{Helmcke}}, \bibnamefont{and}
  \bibinfo{author}{\bibfnamefont{C.}~\bibnamefont{Borde}},
  \bibinfo{journal}{Phys. Rev. Lett.} \textbf{\bibinfo{volume}{67}},
  \bibinfo{pages}{177} (\bibinfo{year}{1991}).

\bibitem[{\citenamefont{Tarallo et~al.}(2014)\citenamefont{Tarallo, Mazzoni,
  Poli, Zhang, Sutyrin, and Tino}}]{tara14}
\bibinfo{author}{\bibfnamefont{M.}~\bibnamefont{Tarallo}},
  \bibinfo{author}{\bibfnamefont{T.}~\bibnamefont{Mazzoni}},
  \bibinfo{author}{\bibfnamefont{N.}~\bibnamefont{Poli}},
  \bibinfo{author}{\bibfnamefont{X.}~\bibnamefont{Zhang}},
  \bibinfo{author}{\bibfnamefont{D.}~\bibnamefont{Sutyrin}}, \bibnamefont{and}
  \bibinfo{author}{\bibfnamefont{G.}~\bibnamefont{Tino}},
  \bibinfo{journal}{Phys. Rev. Lett.} \textbf{\bibinfo{volume}{113}},
  \bibinfo{pages}{023005} (\bibinfo{year}{2014}).

\bibitem[{\citenamefont{Hu et~al.}(2017)\citenamefont{Hu, Poli, Salvi, and
  Tino}}]{hupo17}
\bibinfo{author}{\bibfnamefont{L.}~\bibnamefont{Hu}},
  \bibinfo{author}{\bibfnamefont{N.}~\bibnamefont{Poli}},
  \bibinfo{author}{\bibfnamefont{L.}~\bibnamefont{Salvi}}, \bibnamefont{and}
  \bibinfo{author}{\bibfnamefont{G.}~\bibnamefont{Tino}},
  \bibinfo{journal}{Phys. Rev. Lett.} \textbf{\bibinfo{volume}{119}},
  \bibinfo{pages}{263601} (\bibinfo{year}{2017}).

\bibitem[{\citenamefont{del Aguila et~al.}(2017)\citenamefont{delAguila,
  Mazzoni, Hu, Salvi, Tino, and Poli}}]{agui17}
\bibinfo{author}{\bibfnamefont{R.}~\bibnamefont{del Aguila}},
  \bibinfo{author}{\bibfnamefont{T.}~\bibnamefont{Mazzoni}},
  \bibinfo{author}{\bibfnamefont{L.}~\bibnamefont{Hu}},
  \bibinfo{author}{\bibfnamefont{L.}~\bibnamefont{Salvi}},
  \bibinfo{author}{\bibfnamefont{G.}~\bibnamefont{Tino}}, \bibnamefont{and}
  \bibinfo{author}{\bibfnamefont{N.}~\bibnamefont{Poli}},
  \bibinfo{journal}{arXiv:1712.01388}  (\bibinfo{year}{2017}).

\bibitem[{\citenamefont{Graham et~al.}(2013)\citenamefont{Graham, Hogan,
  Kasevich, and Rajendran}}]{grah13}
\bibinfo{author}{\bibfnamefont{P.}~\bibnamefont{Graham}},
  \bibinfo{author}{\bibfnamefont{J.}~\bibnamefont{Hogan}},
  \bibinfo{author}{\bibfnamefont{M.}~\bibnamefont{Kasevich}}, \bibnamefont{and}
  \bibinfo{author}{\bibfnamefont{S.}~\bibnamefont{Rajendran}},
  \bibinfo{journal}{Phys. Rev. Lett.} \textbf{\bibinfo{volume}{110}},
  \bibinfo{pages}{171102} (\bibinfo{year}{2013}).

\bibitem[{\citenamefont{Hartwig et~al.}(2015)\citenamefont{Hartwig, Abend,
  Schubert, Schlippert, Ahlers, Posso-Trujillo, Gaaloul, Ertmer, and
  Rasel}}]{hart15}
\bibinfo{author}{\bibfnamefont{J.}~\bibnamefont{Hartwig}},
  \bibinfo{author}{\bibfnamefont{S.}~\bibnamefont{Abend}},
  \bibinfo{author}{\bibfnamefont{S.}~\bibnamefont{Schubert}},
  \bibinfo{author}{\bibfnamefont{D.}~\bibnamefont{Schlippert}},
  \bibinfo{author}{\bibfnamefont{H.}~\bibnamefont{Ahlers}},
  \bibinfo{author}{\bibfnamefont{K.}~\bibnamefont{Posso-Trujillo}},
  \bibinfo{author}{\bibfnamefont{N.}~\bibnamefont{Gaaloul}},
  \bibinfo{author}{\bibfnamefont{W.}~\bibnamefont{Ertmer}}, \bibnamefont{and}
  \bibinfo{author}{\bibfnamefont{E.}~\bibnamefont{Rasel}},
  \bibinfo{journal}{New J. Phys.} \textbf{\bibinfo{volume}{17}},
  \bibinfo{pages}{035011} (\bibinfo{year}{2015}).

\bibitem[{\citenamefont{Norcia et~al.}(2017)\citenamefont{Norcia, Cline, and
  Thompson}}]{norc17}
\bibinfo{author}{\bibfnamefont{M.}~\bibnamefont{Norcia}},
  \bibinfo{author}{\bibfnamefont{J.}~\bibnamefont{Cline}}, \bibnamefont{and}
  \bibinfo{author}{\bibfnamefont{J.}~\bibnamefont{Thompson}},
  \bibinfo{journal}{Phys. Rev. A.} \textbf{\bibinfo{volume}{96}},
  \bibinfo{pages}{042118} (\bibinfo{year}{2017}).

\end{thebibliography}
\end{document}